# Piezo-generated charge mapping revealed through Direct Piezoelectric Force Microscopy

A.Gomez[a]*, M. Gich[a], A. Carretero-Genevrier[bc], T. Puig[a], X. Obrador[a]

[a]Institut de Ciència de Materials de Barcelona (ICMAB-CSIC), Campus UAB, Bellaterra, 08193, Catalonia, Spain
[b]Institut des Nanotechnologies de Lyon (INL) CNRS-Ecole Centrale de Lyon, 36 avenue Guy de Collongue, 69134 Ecully, France
[c]Institut d'Electronique et des Systemes (IES), CNRS, Universite Montpellier 2 860 Rue de Saint Priest 34095 Montpellier

ABSTRACT

While piezoelectrics and ferroelectrics are playing a key role in many everyday applications, there are still a number of open questions related to the physics of those materials. In order to foster the understanding of piezoelectrics and ferroelectric and pave the way to future applications, the nanoscale characterization of these materials is essential. In this light, we have developed a novel AFM based mode that obtains a direct quantitative analysis of the piezoelectric coefficient $d_{33}$. This nanoscale tool is capable of detecting and reveal piezo-charge generation through the direct piezoelectric effect at the surface of the piezoelectric and ferroelectric materials. We report the first nanoscale images of the charge generated in a thick single crystal of Periodically Poled Lithium Niobate (PPLN) and a Bismuth Ferrite ($BiFO_3$) thin film by applying a force and recording the current produced by the materials. The quantification of both $d_{33}$ coefficients for PPLN and BFO are 13 ± 2 pC/N and 46 ± 7 pC/N respectively, in agreement with the values reported in the literature. This new mode can operate simultaneously with PFM mode providing a powerful tool for the electromechanical and piezo-charge generation characterization of ferroelectric and piezoelectric materials.

INTRODUCTION

The piezoelectric effect, which consists in the dielectric polarization of non-centrosymmetric crystals under a mechanical stress, was discovered by the Curie brothers in 1880[1]. The following year, from thermodynamic considerations, G. Lippmann predicted the converse effect, *i.e.* that a piezoelectric material would be mechanically strained by an applied electric field[2] and the Curies readily measured it[3]. These findings spawned more research which eventually led to the discovery of ferroelectricity in polar piezoelectrics[4]. Since those early discoveries, the unique ability of piezoelectrics and ferroelectrics for interconverting mechanical and electrostatic energies[5] has endlessly inspired technological developments and these materials, which represent nowadays a billion euro industry, are found in many everyday applications[6–12]: ultrasound generators for echography scanners, shock detectors within airbags, accelerometers, diesel injection valves, tire pressure sensors, vibration dampers, oscillators, improved capacitors, or new dynamic access random memories, to just cite a few. Moreover, the prospects for future applications in new markets are bright, including energy harvesting, CMOS replacement switches, or photovoltaics and photocatalysis[13–16]. Yet, in spite of such industrial relevance and the amount of past and present research, the basic understanding of piezoelectricity and ferroelectricity is challenged and reshaped by findings that come along with new developments in the characterization of materials. This is well illustrated by the advances in atomic force microscopy, which brought a new perspective of ferroelectric domain walls[17–20]. The development of new modes and an improved spatial resolution have revealed the domain wall complexity and its intrinsic properties [21–23] and have also opened the door to get more insight in long-date issues such as the extrinsic contributions to dielectric permittivity and piezoelectricity due to domain wall pinning at dislocations and grain[24]. In this direction, Piezoresponse Force Microscopy (PFM) is the most widely used technique for the nanoscale and mesoscale characterization of ferroelectric and piezoelectric materials[25–28]. PFM method is based on the converse piezoelectric effect and consists in measuring the material deformation under an AC electric field applied through the contacting AFM tip. In this technique the sample vibration is determined by an optical beam deflection system, which is an indirect measurement[29], making the accurate determination of the piezoelectric coefficient challenging. Moreover, the quantitative piezoelectric measurements by PFM[30], are further complicated by the difficulty of disentangling, from the electromechanical response, the contributions of the piezoelectric response and other physical phenomena such as, ionic motion and charging, electrostatic or thermal effects[18,31–33]. Indeed, the increasing awareness about these issues among the scientists of the field[34] prompts the need for new developments in scanning probe microscopies, which remain a unique tool for the characterization of piezoelectric and ferroelectric materials at the nanoscale.

To address this need, here we introduce a new SPM tool that exploits the direct piezoelectric effect to obtain a quantitative measurement of the piezoelectric constant in piezoelectrics. This technique, that we call Direct Piezoelectric Force Microscopy (DPFM) uses a specific amplifier and a conductive tip which simultaneously strains a piezoelectric material and collects the charge built up by the direct piezoelectric effect. The amplifier is an ultralow input bias current (<0.1 fA) transimpedance capable of converting electric charge into a voltage signal, readable by any commercial microscope (see figure 1a). As a consequence, the developed setup has a very low leakage current, and thus all the charges generated by the piezoelectric material can be read by the amplifier. Just by maintaining the tip on the surface of the films and sequentially applying different force values with the AFM tip, the charges generated by the material are measured and the direct piezoelectric coefficient can be readily calculated from the applied stress and the collected compensation charge. Interestingly, by combining this tool withPFM measurements, a complete electromechanical and piezo-charge generation characterization can be achieved.

Measuring the direct piezoelectric effect with an AFM is a challenge that has not been addressed so far due to the impossibility of performing reliable measurements of tiny amounts of generated charge. An AFM probe can apply a user predefined force with picoNewton precision, up to maximum values of hundreds of microNewtons[35–37]. Applied to a piezoelectric material, such force will generate a charge, which can be collected to obtain currents of different intensity depending on the sampling time. For instance, we can estimate that the 1fC charge generated by applying a 100 μN force into a 10 pC/N piezoelectric material[38], will produce a current of 1 fA if generated in 1 s, 2 fA if generated in 0.5 s and so on. With such requirements, an amplifier capable of measuring 1fA with a BandWidth (BW) of 1 Hz is needed. More importantly, the charge that the amplifier leaks has to be well below that desired threshold of 1fA, otherwise a substantial part of the current will be lost during measurements. Since these requirements were not met by any AFM manufacturing companies, a special amplifier was employed.

EXPERIMENTAL SETUP

The complete setup to perform measurements according to the proposed method is depicted in **Figure 1a**. The amplifier consists of three different commercially available Operational Amplifiers (OA), which were supplied by Analog Devices Inc. The amplification process is divided into two stages, a transimpedance stage and a voltage amplifier stage. The transimpedance stage was configured with a feedback resistor of 1TeraOhm which yields a Current-to-Voltage gain of -1x10$^{12}$ V/A[39]. The voltage amplifier stage adds an additional gain of 72,25. Following standard amplifier theory, the final gain of both concatenated stages is the multiplication of each stage gain, which results in a gain of -72,25 x 10$^{12}$ V/A[40]. Even though theoretical gain calculation is precise, we experimentally calibrated the amplifier twice with a test resistor of 40 ± 0,4 GOhm giving an experimental gain of -16,9 ± 1,0 x 10$^{12}$ V/A (see Figure S1 of Supplementary Information). The leakage current through the amplifier induces an error, which will be responsible of charge losses while measuring. Such current was provided by Analog Devices as being as low as 0,1 fA, which can be considered small compared to the generated piezocharge[41] to be measured, which is in the order of several fA. An intrinsic property of the setup is that both tip and back-surface of the sample are connected to ground, which enables the study of high leakage ferroelectric films.

With such setup the charge generated by a piezoelectric material can be recorded with an AFM tip. The physics underlying the generated current is depicted in **Figure 1b, c and d**. Two different cases are considered, when the tip scans from left to right (Trace) and from right to left (Retrace). While in trace scanning, **Fig. 1b**, the moving tip creates a strained area on the right side of the tip apex, while the area on its left side is unstrained. When an up domain polarization is scanned, a positive charge (+Q) is generated in the strained region implying a positive flowing current. In contrast, a charge of opposite sign is created (-Q) in the unstrained area on the left side of tip apex. The charges generated at the strained and unstrained regions cancel out, yielding a zero net charge, because the strained and unstrained charge generation processes are compensated. Nevertheless, the situation is completely different at domain walls. Once the tip apex is located on the domain wall, the strained region, which now has a downwards polarization, will generate a negative charge (-Q). The unstrained up polarization region will remain unchanged, generating a negative charge (-Q). Thus, a negative charge is generated, which can be quantified by measuring the negative current flowing through the tip. In this case, the measured current corresponds to the tip loading the down polarization and unloading the up polarization. Similarly, when the tip scans from right to left the unstrained region corresponds now to the down polarization state, and hence, a positive charge (+Q) is generated (see **Fig 1c**). At the same time, the strained region, which is in the up polarization state, will generate a positive charge (+Q). As a result, a positive charge is generated at the domain wall and a positive current can be measured by the AFM tip. Again, no net charge results from scanning a single domain, as the strained and unstrained regions will generate charges of opposite signs. Spectroscopy experiments can also be performed, see **Fig 1d**, as the tip exerting a force generates a positive charge (+Q), if an up domain is loaded, or a negative charge (-Q) if a down domain is loaded. By the contrary, the unloading process generates a negative charge (-Q) for an up domain and a positive charge (+Q) for a down domain. As current is being recorded, the rate at which the force is applied rate is crucial, as the current increases with force rate. Throughout the manuscript it is considered that a positive force-straining force- will generate a positive current if applied into a positive (up) poled domain.

In the experiments we used a commercial probe with reference RMN-25PT200H. The tip is made out of a solid platinum wire consisting in an ultra stiff cantilever, with spring constant of 250 N/m. Such fully metallic tip ensures that its conductivity nature is preserved while applying a high load and only a decrease in resolution can eventually occur. We tested the new mode on a typical reference material for PFM experiments which is a commercially available Periodically Poled Lithium Niobate (PPLN)[42] in the form of a thick crystal. This material has been widely studied and its $d_{33}$ piezoelectric constant is in the range of 6-16 pC/N[43]. Before starting the measurements, the sample was scanned with the conductive tip in order to discharge its surface from screening charges and minimize their effects[44,45].

RESULTS

Through the aforementioned setup and the proposed physical explanation, we have been able to perform the first mapping of piezoelectricity at the nanoscale. The output signal of the amplifier was both recorded at the Trace (**Figure 2a**) and Retrace (**Figure 2b**) scans. The images consist of a 256x128 pixels frames, 15 µm x 30 µm obtained at a speed of 0.01 lines/s (ln/s) (0,66 µm/s) and were recorded with a loading force of 234 µN. We used a particularly low speed to avoid scrapping surface screening charge which could interfere with the collected charge[46]. With these imaging parameters bandwidth needed to record current is 5 Hz, which is in accordance with what our amplifier can perform. The obtained images (see **Figure 2a,b**), show that the current is only recorded at the domain walls in accordance with the proposed physical model. A peak current of 15 fA is generated at the domain walls while its sign depends on the direction of the tip scan. We labeled the "Trace" image, from left to right, as "DPFM-Si", for Direct PFM Signal input, and the "Retrace" image, from right to left, as "DPFM-So", for Direct PFM Signal output. We have also tried to perform both PFM and DPFM methods, simultaneously. In order to do so, the back of the PPLN crystal was connected to the AC generator of the AFM, so an AC voltage signal was applied to the bottom surface of the sample maintaining a DC coupled ground. The PFM phase image is shown in **Figure 2c** and PFM amplitude image is shown in **Figure 2d**. The simultaneous acquisition of the four images of Figure 2 shows how the DPFM mode can complement the standard PFM measurements providing, as we will discuss below, the data to quantify the piezoelectric coefficient of the material. Moreover, standard topography image and friction image, obtained from contact mode operation are recorded (see Figure S2 of SI). From DPFM-Si and DPFM-So images it is observed that there is a little gradient in the single domains areas, this will imply the collected current is not exactly zero. This could be due to different processes occurring simultaneously with piezoelectric charge generation as, for instance, surface screening recharging[47]. However its contribution is negligible compared to the peaks recorded at domains walls (see Figure S3 of SI).

In order to obtain strong evidence of the piezoelectric origin of the current signal from the amplifier we prepared a full set of experiments related to the dynamics of piezoelectric charge generation. The charge generated from piezoelectric effect is known to be linear with the applied force[5]. This is a key aspect to distinguish piezoelectric charge from other possible charge generation phenomena[46,48]. The relationship between current and applied load was tested by scanning the PPLN sample under different applied loads, starting from a low loading force of 9 µN which was stepwise increased until reaching a maximum force of 234 µN. The recorded DPFM-Si and DPFM-So images are plotted in **Figure 3a** and **Figure 3b**, respectively. The tip speed was maintained constant along the whole image at a rate of 0,55 µm/s. We can observe that at the lowest load, no charge was collected by the amplifier, which was not capable of read such a small current, i.e between 0,1-0,3 fA for an applied force of 9 µN. The area recorded with the minimum force loading is also interesting to assess the influence of surface charge screening in the recorded currents. Before DPFM experiments, the sample was scanned with the same tip, at a tip speed 100 times faster, in order to fully discharge the sample surface from surface screening charge. The area scanned with 9 µN confirms that surface screening charges do not play an important role in the collected charge. If removal of surface screening charge through a scrapping process was important we should see a current in the 9 µN region, as the applied force is two-fold that needed to start the scrapping process[44]. Once the force is increased, the current recorded by the amplifier increases as well, as it should be expected from a piezoelectric generated charge. More importantly, the width of the current line generated at domain walls does not substantially increase with applied load. The size of this line is not related to the domain wall thickness, but to a convolution effect caused by the tip[49,50] (see S4 in SI).

In order to elucidate if the generated charge is proportional to the force we have analyzed the peak current values for DPFM-Si and DPFM-So frames, for each applied load. The maximum current values of a scan line were multiplied by the specific time constant of one pixel, which is 0,39 s, so the most part of the piezoelectric charge is fully integrated. Finally, a relation between the collected Charge vs Applied load is found, which is plotted in **Figure 3c**. A linear fit was used for both positive and negative charge generated confirming the linear relationship between the generated charge and the applied force with Pearson's R of 0,99 and -0,93 for each linear fitting. From the slope of this linear fit, an approximation of the $d_{33}$ piezoelectric constant of the material can be found with a value of 8,2 pC/N. The value obtained is an underrated approximation, as there is a part of the current generated that it is not being considered, as only the peak current is integrated. The current profile shape for each applied load was also analyzed, which are plotted in **Figure 3d**. The profiles provide information on the dynamics of the charge generation at the nanoscale as the tip passes throughout the domain wall. It is found that the piezoelectric current has a Gaussian-like shape, where the area below the Gaussian curve is the piezo-generated charge. The profiles, evidence that the increased generated charge for higher loads is related to the maximum current peak, rather than to the width of the Gaussian-like curve shape. This is in accordance with the fact that the tip does not significantly increase its radius with the applied load. Once the origin of the generated charge has been proved to be the direct piezoelectric effect, we can now perform a mapping of the piezopower generation at the nanoscale with images of Figure 2 (see S5 in SI).

Obtaining quantitative values of piezoelectric and ferroelectric materials through an easy and reliable method is a high pursued target in the scientific community[51,52]. In order to test if the method can be quantitative, we performed a zoomed-in image of a domain wall, recording both DPFM-Si and DPFM-So signals, see **Figure 4a** and **Figure 4b**. The images were performed with a tip speed of 0,22 µm/s and an applied load of 234 µN. The zoomed in images were sufficiently precise to fully integrate the generated current. In order to reduce thermal noise[53], the mean average profile for the total number of lines composing the image was obtained for both cases, see **Figure 4c** and **Figure 4d**. The resulting profile corresponds to the piezoelectric generated charge vs distance (µm) which divided by the tip velocity value can be converted into charge vs time. With such experimental profiles, see **Figure 4c and 4d**, we can perform a gaussamp fit of the obtained curves to estimate the area beneath the curve. We have found that the piezoelectric charge generated is 5,7 ± 0,4 fC for DPFM-Si and 6,5 ± 0,5 fC for the DPFM-So profiles. In order to see if the collected charge is a function of the tip speed we studied the evolution of the recorded charge versus tip speed (see S6 in SI). The measured charge corresponds to a loading and unloading mechanism, and hence to find the piezoelectric charge we must divide this charge by a factor of two. The exact force exerted was calculated using a Force-vs-Distance curve, (see S7 in SI) and with such deflection sensitivity and the cantilever spring constant, the applied force was obtained. To diminish the error associated to the applied force, we have calculated the exact force constant of the probe used in the experiment, through a formula provided by the tip manufacturer and the real dimensions of the cantilever. Upon calculations, we found that the applied load is 234 µN, which yields a piezoelectric constant of 12,1 pC/N and 13,8 pC/N, for DPFM-Si and DPFM-So, respectively. This is in accordance with the value found in the bibliography, where the $d_{33}$ constant of PPLN is in the order of 6-16 pC/N. In fact, we have evaluated the error that corresponds to the proposed method. The force error was found to be ± 9 µN, mainly caused by the determination of the spring constant of the cantilever. The charge measurement error was calculated as the sum of the statistical error, the error created from the amplifier leakage current and the error obtained from the electrical calibration. The total error is ± 0,7 fC for DPFM-Si and ± 0,9 fC for DPFM-So profiles. Summing all the errors, we found that the $d_{33}$ piezoelectric constant of our sample is 12 ± 3 pC/N and 14 ± 4 pC/N for DPFM-Si and DPFM-So respectively. As we are crossing the very same domain, we can use both quantities to acquire the final $d_{33}$ constant of the material as being 12,9 ± 2,4 pC/N standard error.

Spectroscopy experiments were performed to elucidate if the method could also be employed not only for imaging, but also as a tool of characterizing the piezoelectric response outside the ferroelectric domain walls or in non-ferroelectric piezoelectrics. For such purpose, the tip was placed in the middle of a ferroelectric domain and the current recorded while a Force-vs-Distance curve was obtained. The curve starts with a loading force of 5 µN and it is increased to a maximum value of 258 µN to go back to the initial 5 µN load. The current recorded from the amplifier was measured for different applied force rates, see **Figure 4e**. As the force/time rate is increased, the recorded current increases as well confirming its direct relationship. Different spectroscopy events were obtained, see **Figure 4f**; top which corresponds to a spectroscopy for up domain area and bottom for down domain area. It is found that for the up domain case, a loading curve will generate a positive current; however the current sign is the opposite in the case of a down polarization domain. The spectroscopy curves started with the tip in contact with the surface under an applied load of 5µN, to avoid collecting charge generated by electrostatic effects while the tip is moved from air to the sample surface. For both curves a force/time ratio of 53,2 µN/s was employed.

The feasibility of the method has been successfully demonstrated for a thick ferroelectric crystal with a low-intermediate piezoelectric $d_{33}$ constant. In order to check if the performance of the DPFM method on other materials it was also tested on a 400nm-thick BFO ferroelectric layer over platinum, commercially available from MTI Corp. The sample was previously scanned using PFM in order to record a pattern in its surface-the pattern is shown in PFM phase image of **Figure 5a**, where DC voltages of +45 VDC and -45VDC were applied to the bottom contact of the sample in order to poll the domains. The same area was scanned using normal PFM mode in order to see if the domains can be read. Once recorded, DPFM-Si and DPFM-So images were performed, which are shown in **Figure 5b** and **Figure 5c**. It is found that the current generated is only present at domain walls, however with a similar scanning parameters, it is found that the peak current is near 25fA. BFO is a well-characterized ferroelectric that has a young modulus of 170 GPa and a surface screen charge of 80 µC/cm$^2$[54]. These values are comparable to those of the previously tested PPLN[43]. However, the piezoelectric constant of BFO is significantly larger, between 16-60 pC/N[54]. These differences in the measured $d_{33}$ constants can be used to explain the larger current that is recorded for BFO compared to PPLN. In order to discard imaging artifacts, the same pattern was reread in DPFM mode but rotating the scan direction, which rotates the image motives as well (see S8 in SI). It was found that the generated charge had its maximum value where the tip passes from a full polarized area to the opposite polarization direction. The capability of the mode to be quantitative was again tested by determining the $d_{33}$ value for the BFO sample. The same procedure as explained for **Figure 4c** and **Figure 4d** were employed for **Figure 5b** and **Figure 5c**. The squared area in **5b** and **5c** were used to obtain an average of the lines composing squares resulting in the average profile of **Figure 5d**. The top part corresponds to the A square and the bottom part corresponds to the B square. The profiles were fitted with a gauss-amp curves, red and blue lines respectively. The values obtained for the fitting curves are 24,1 ± 1,7 fC and -26,6 ± 3,4 fC, which divided by the applied force, yield a $d_{33}$ values of 44,1 ± 6,9 pC/N and 48,7 ± 12,7 pC/N for DPFM-Si and DPFM-So profiles, which, averaged, result in a $d_{33}$ value of 46,4 ± 7,2 pC/N. Such value is accordance with what is found in the literature, which ranges between 16 and 60 pC/N[54], confirming the feasibility of the mode as a tool to quantify the piezoelectric coefficient.

CONCLUSIONS

The measurement of charges generated by the direct piezoelectric effect with nanoscale resolution has been demonstrated through the use of a novel AFM based method. The new mode, which we call

Direct Piezoelectric Force Microscopy (DPFM), is based in the direct piezoelectric measurement principle where the piezogenerated charge is collected by applying forces in the µN range to a piezoelectric sample with a conductive AFM tip. We studied the feasibility of this new mode by exploring the piezogenerated charge dynamics of Periodically Poled Lithium Niobate and Bismuth Ferrite ferroelectrics. The simultaneous acquisition of DPFM and standard Piezoresponse Force Microscopy images, can provide a new tool for a better understanding of the electromechanical and piezocharge generation dynamics at the nanoscale. The method was also applied in spectroscopy experiments which allow the determination of the piezoelectric response outside the domain walls and in non-ferroelectrics. We have demonstrated that the new mode is quantitative by measuring the d33 constants for PPLN and BTO, which were in accordance with the valuespreviously reported. The specific nature of AFM, with its high force precision, plus the use of an ultra-low-leakage high precision amplifier makes this mode a promising tool as an accurate, fast and reliable ferroelectric material characterization technique.

ACKNOWLEDGMENTS

ICMAB acknowledges financial support from the Spanish Ministry of Economy and Competitiveness, through the "Severo Ochoa" Programme for Centres of Excellence in R&D (SEV- 2015-0496), and the projects Consolider NANOSELECT (CSD 2007-00041). and MAT2014-51778-C2-1-R project, co-financed with FEDER, as well as the Generalitat de Catalunya (project 2014SGR213).The authors thank ICMAB Scientific and Technical Services. We thank Oliver Anderson from Rocky Mountain Nanotechnology LLC for discussion on how to calibrate cantilever spring constant.

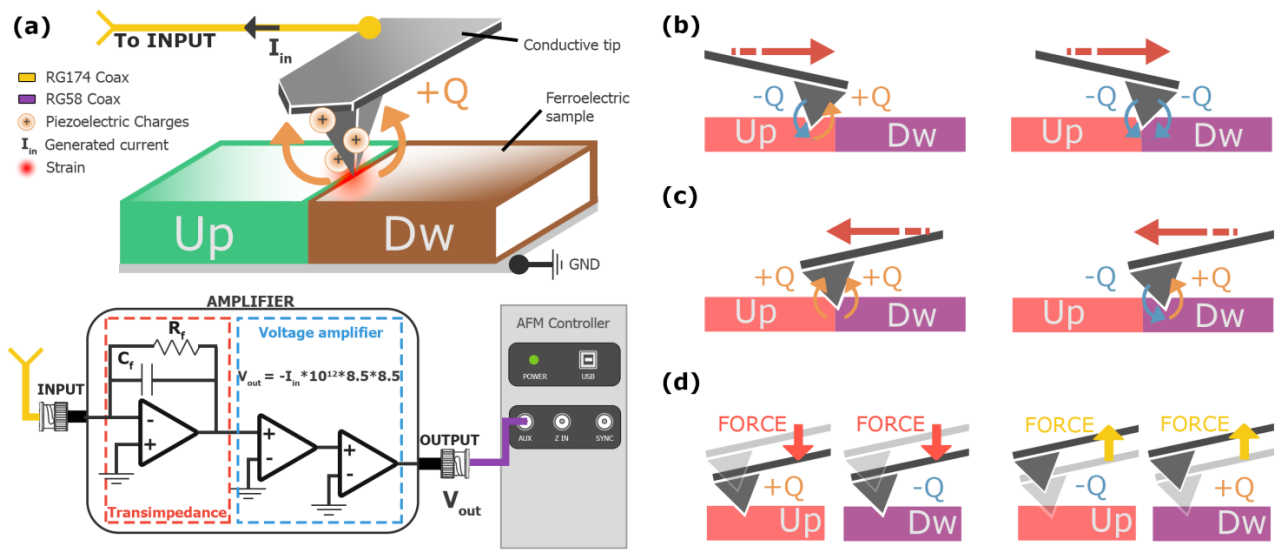

**Figure 1a,** Setup used to record the piezoelectric charge generated by the material through the use of a special current-to-voltage transimpedance amplifier. The amplifier maintains a reasonable bandwidth of 4-5 Hz with an ultra low input-bias current consumption of less than 0.1 fA. **Fig. 1b,** Qualitative model explaining the charge generated in a ferroelectric material during the scan of an AFM tip in contact mode. As the tip moves from left to right, material is strained at the right side of the tip apex and is in an unstrained state at the left part of tip apex. While the tip scans a single domain, the strained and unstrained regions generate charges of opposite signs and hence the net current is zero. When the tip scans the domain wall region, the generated charge present the same sign and hence a piezogenerated charge is created. **Fig. 1c,** Physical model of the charge generated sign once the tip scans from right to left. The strained region is located at the left part of the tip apex, while the unstrained region is at the right side. Charge generation occurs again at domain walls, but with an sign opposite to that of a right-left scan. **Fig 1d**, Spectroscopy sweep model obtained when the tip performs a Force-vs-distance sweep. While the tip exerts a force on the sample a strain is created. Once the force is released, "unstrain" occurs. The strain and unstrain processes create positive or negative charges generated depending on the polarization of the domain.

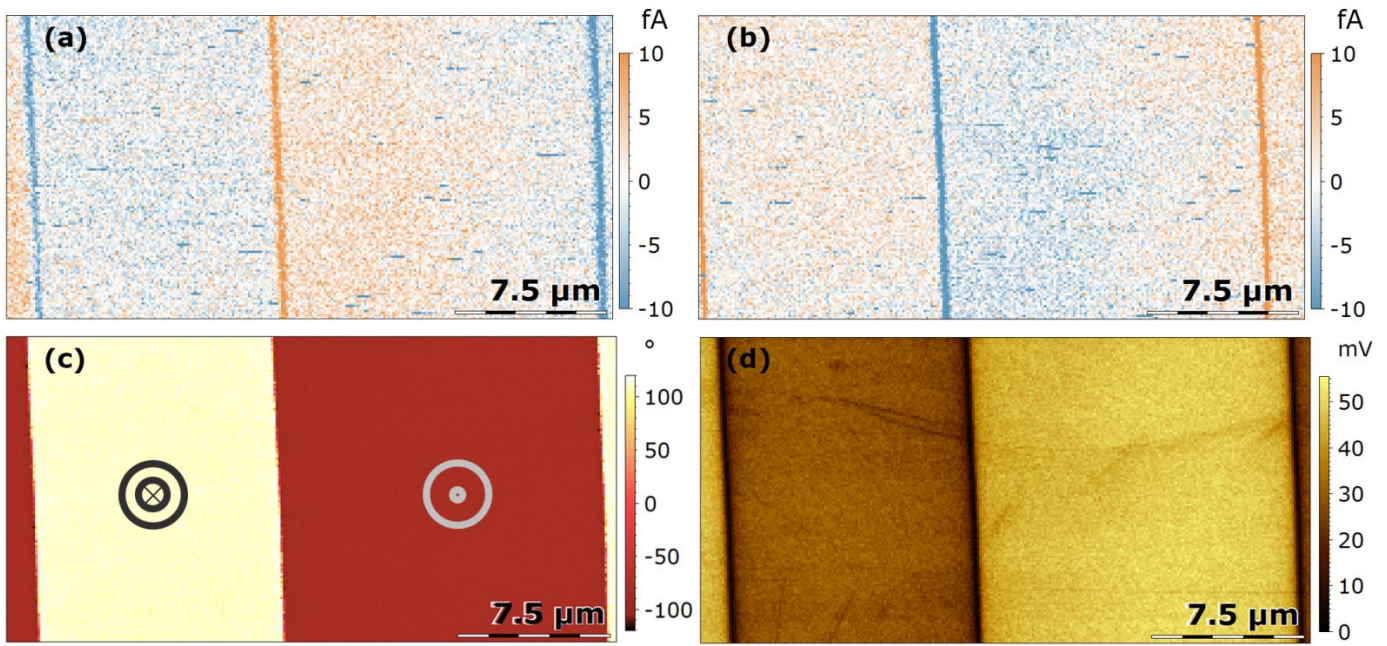

**Figure 2a** Piezo-generated current map obtained when the tip scans from left to right-trace (DPFM-Si). **Fig 2b** Piezo-generated current map obtained when the tip scans from right to left-retrace (DPFM-So). Current is generated at domain walls, orange and blue vertical lines, where the tip strains and unstrains opposite domains. Inside the domains, a near zero current signal is observed, however a little contrast is present that can be due to surface screening recharging process. **Fig 2c** PFM phase image and **Fig 2d** PFM amplitude image of the same sample, obtained simultaneously with DPFM signals. In order to obtain DPFM signals an AC bias was connected to the back electrode of the specimen.

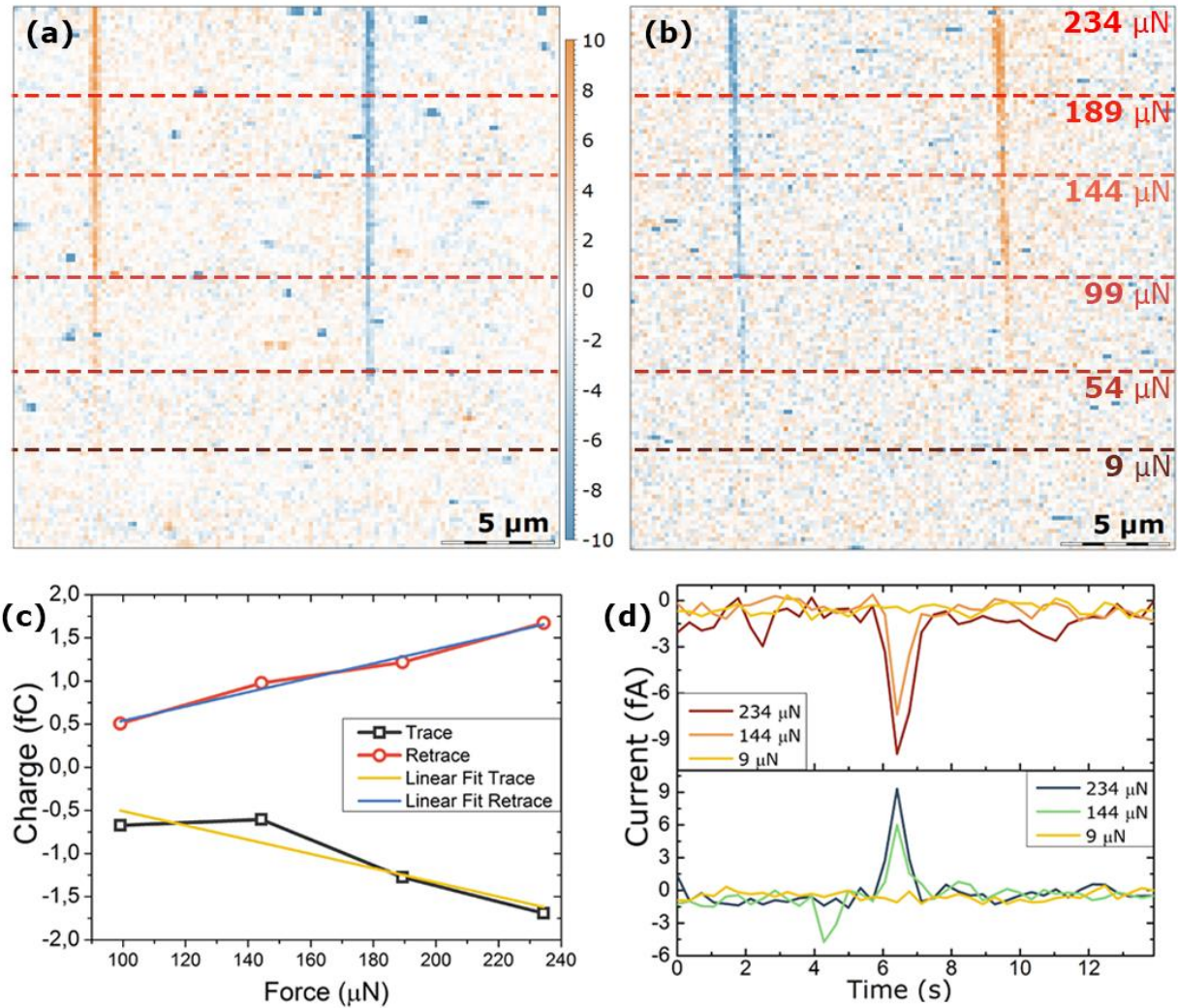

**Figure 3a** DPFM-Si and **Figure 3b** DPFM-So of the proposed PPLN test sample, obtained at different applied forces. In order to demonstrate the origin of the recorded current, different forces where applied during the scan-see red line dot. The current recorded increases with the applied force, as expected from a piezoelectric charge generation. **Figure 3c** Charge vs Force spectroscopy sweep obtained from the profiles of Figure 3a. The current profiles where integrated with a time constant of 390 ms in order to obtaine the charge generated at a specific pixel. The linear relation displayed between force and charge collected confirms the piezoelectric nature of the generated charge. **Figure 3d** current profiles extracted from Fig. 3a for different applied forces. Applying 9 µN is not enough to read the current generated as it lies below the current threshold of the amplifier. As the force is increased, the amplifier responds to the generated charge, within a symmetric gaussamp like curve.

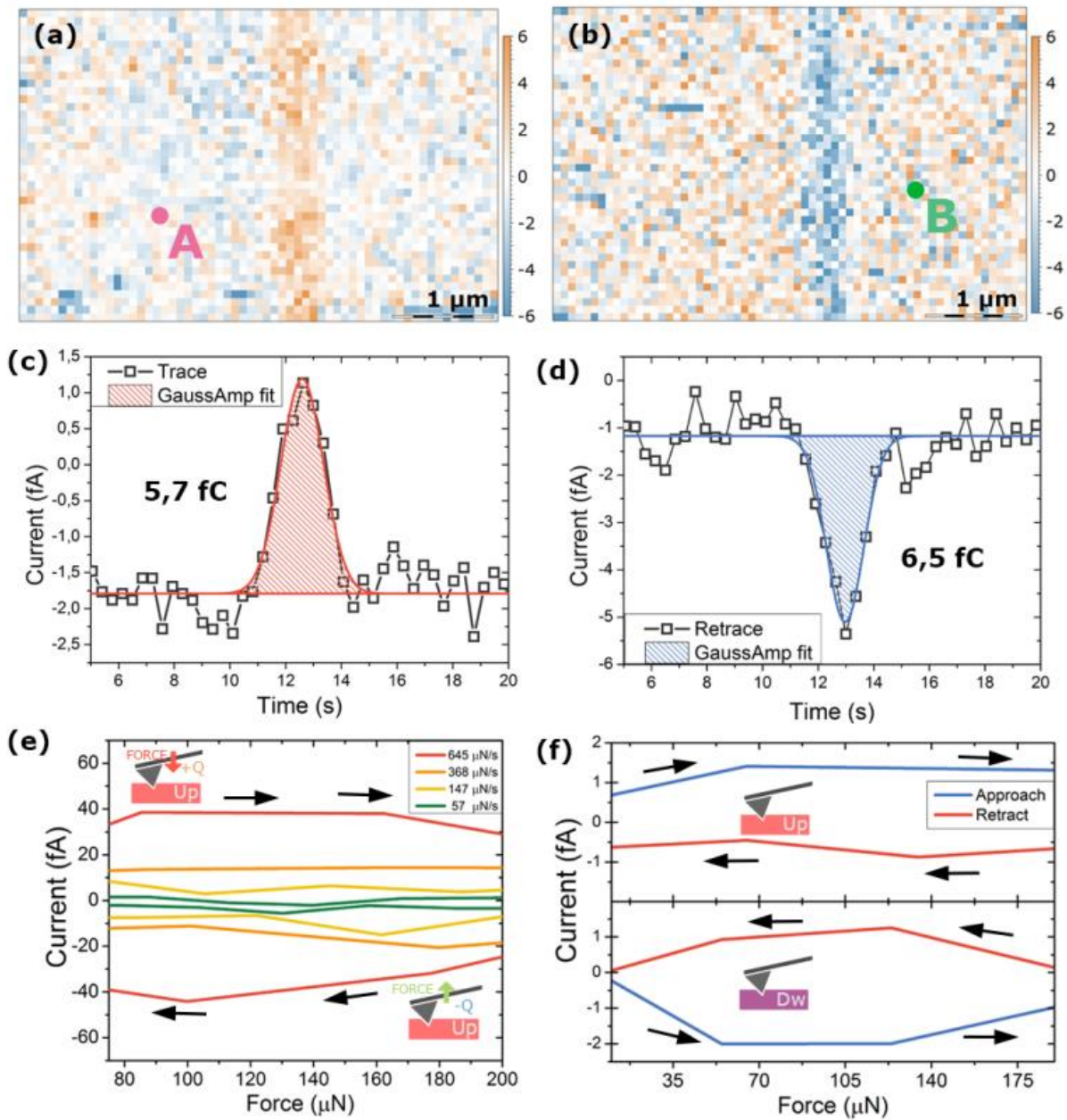

**Figure 4a** DPFM-Si and 4b DPFM-So images of a zoomed region of the PPLN sample recorded in order to fully integrate the charge generated. The mean profile average from the images was obtained in order to reduce noise. The resulting profiles are plotted in **Figure 4c** and **Figure 4d**. A symmetric gauss-amp fitting curve was performed in order to estimate the charge production of the PPLN material as the integral of the fitting curve (crossed filled area). **Figure 4e** Current-vs-Force spectroscopy sweep performed in the Up domain configuration, where different Force sweep rates where applied. The current generated increases with the increasing force rate. Its sign is the opposite for approach-when force increases- and retract-when force decreases. **Figure 4f** Current vs Force spectroscopy sweep for an Up domain (top) and a Down domain (bottom). For an Up domain increasing the force will generate a positive current while the opposite occurs for a Down domain.

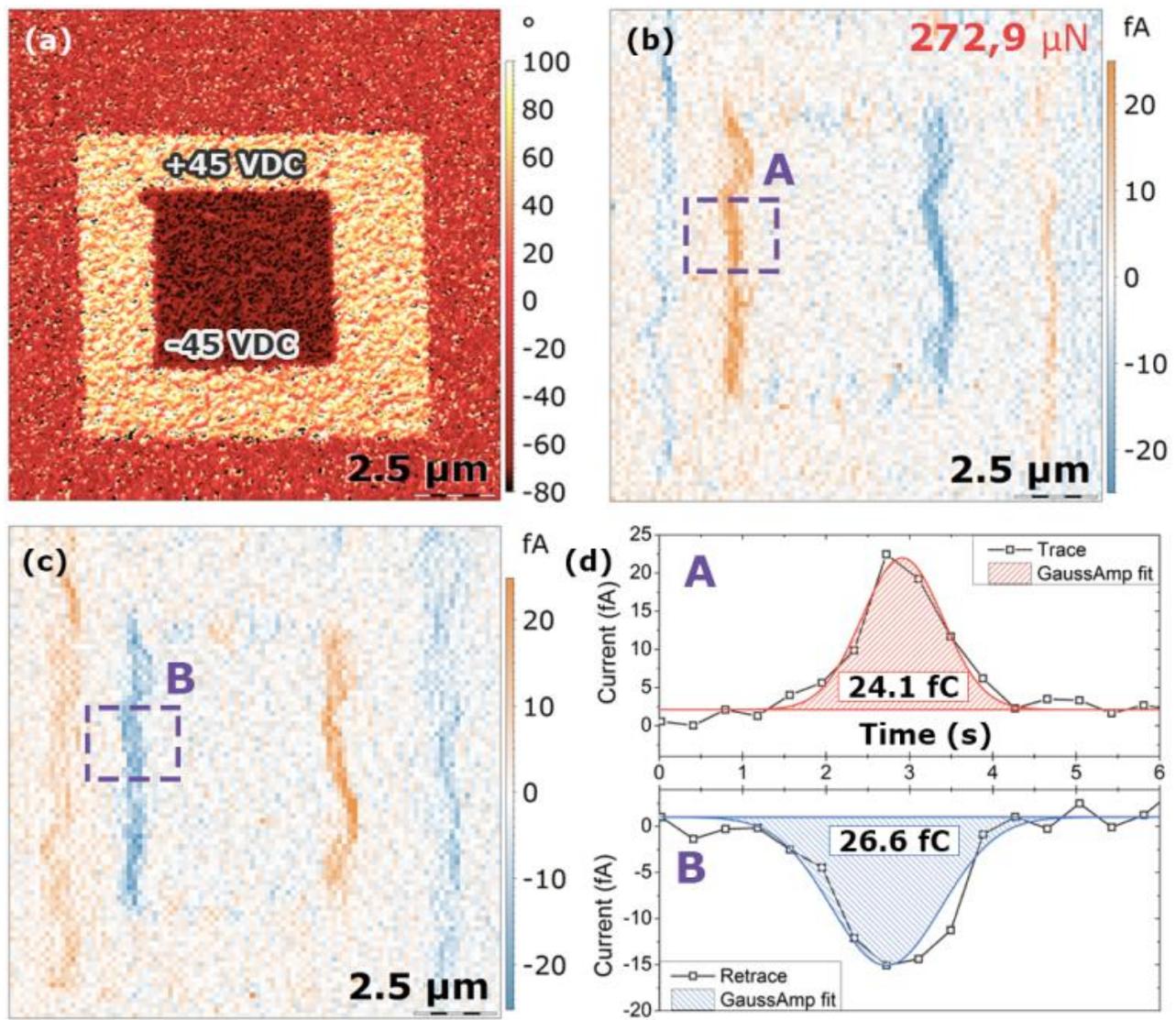

**Figure 5a** PFM phase image obtained in a prerecorded 400 nm thick BiFeO$_3$ (BFO) ferroelectric sample. **Figure 5b** DPFM-Si image and **Figure 5c** DPFM-So image of the prerecorded area. The inner, smallest square recorded has the highest current output, as we are crossing two distinct fully polarized domains. However, largest recorded square produces less current output, as we are crossing from a virgin state to a fully polarized domain. **Figure 5d** Mean-profile of the dot-line squared area A- in Figure 5b- top and Mean profile of the dot-line square B of Figure 5c. Both profiles where fitted with a GaussAmp curve in order to obtain the charge generated by the material and hence its piezoelectric constant.